# B-BOP, the SPICA Imaging Polarimeter

V. Revéret[*a], M. Sauvage[a], Obaïd Adami[a], A. Aliane[b], M. Berthé[a], S. Bounissou[a], X. de la Broïse[a], M. Chimeno[d], A. Demonti[a], J. Delabrouille[a,g], C. Delisle[a], E. Doumayrou[a], L. Duband[c], D. Dubreuil[a], L. Dussopt[b], P.-A. Frugier[a], C. Gennet[a], O. Gevin[a], V. Goudon[b], H. Kaya[b], B. Marquet[e], J. Martignac[a], S. Martin[c], P. Najarro[f], X.-F. Navick[a], J. Perez[d], J.-Y. Plesseria[e], A. Poglitsch[a], L. Rodriguez[a], J. Torres Redondo[f], F. Visticot[a]

[a]Université Paris-Saclay, CEA IRFU, Gif-Sur-Yvette, France; [b]CEA-Grenoble, LETI-DOPT, Grenoble France ; [c]CEA-Grenoble, DRT/DSBT, Grenoble, France ; [d]UPM Madrid, Spain, [e]CSL Liège, Belgium, [f]CAB-INTA, Spain, [g]APC, Université de Paris, Bâtiment Condorcet, Paris France

**ABSTRACT**

We present the B-BOP instrument, a polarimetric camera on board the future ESA-JAXA SPICA far-infrared space observatory. B-BOP will allow the study of the magnetic field in various astrophysical environments thanks to its unprecedented ability to measure the linear polarization of the submillimeter light. The maps produced by B-BOP will contain not only information on total power, but also on the degree and the angle of polarization, simultaneously in three spectral bands (70, 200 and 350 microns). The B-BOP detectors are ultra-sensitive silicon bolometers that are intrinsically sensitive to polarization. Their NEP is close to $10E^{-18}$ W/sqrt(Hz). We will present the optical and thermal architectures of the instrument, we will detail the bolometer design and we will show the expected performances of the instrument based on preliminary lab work.

## 1. INTRODUCTION

SPICA[5] (SPace Infrared telescope for Cosmology and Astrophysics), one of three remaining candidates for the ESA Medium Class M5 mission and due to compete for the M5 slot through ESA's Mission Selection Review process starting in early 2021, was unexpectedly cancelled by ESA in October 2020 on financial grounds. With an actively cooled 2.5m-diameter telescope, this space observatory would have comprised three instruments (SAFARI – SpicA FAR-infrared Instrument, SMI – SPICA Mid-Infrared Instrument and B-BOP – magnetic field (B) explorer with BOlometric Polarimeter) working in synergy to provide unprecedented spectroscopic and photometric sensitivity in the mid- and far-infrared, transforming infrared astronomy in fields ranging from planet formation to star formation through to studies of the interstellar medium and galaxy evolution. This paper is one in a series of SPIE contributions from the SPICA Consortium that is aimed at preserving the technological developments and knowledge gained through work undertaken by hundreds of scientists over several years, ensuring that the legacy of SPICA is not lost through the untimely cancellation of this mission.

### 1.1 Science case for B-BOP

In the far-infrared (FIR), the main mechanism for producing linearly polarized light is grain alignment with the magnetic field. Quite a number of unknowns exist regarding the actual mechanism that produces this alignment but the observational basis for it is firm. Therefore, characterizing the level of polarization in the FIR is a method to access the geometric properties of the magnetic field in the interstellar medium (ISM).

Considering that the fraction of energy that is stored in the magnetic field is comparable to that which is present as kinetic or gravitational energy, it is clear that the magnetic field must play a part in all the processes that give rise to the heavily structured ISM we observe and to the key process of star formation.

---

[*] Send correspondence to V. Revéret : vincent.reveret@cea.fr

Therefore, the main objective of B-BOP is to provide, through its capacity to measure the polarized fraction and polarization angle, large scale maps of the magnetic field structure in the ISM. This will be performed in star-forming and non-star-forming regions of our own galaxy, as well as of its close neighbours where the spatial resolution provided by SPICA allows separating ISM clouds.

Moving on to galaxies beyond the Local Group, B-BOP will study how the parsec-scale magnetic field is connected to the large-scale magnetic field in galaxies to shed light on the process, possibly dynamo, that gives rise to it in the course of galaxy evolution. Understanding whether and how the large-scale magnetic field structure is connected with the mechanism that allows Active Galactic Nuclei to affect star formation at the scale of their host galaxies will be another aspect of B-BOP's science objectives in the extragalactic domain.

Finally, the possibility that a polarization signal subsists in the cosmic infrared background is under investigation as it probably contains interesting clues on the processes of galaxy assembly.

## 2. DESCRIPTION OF THE INSTRUMENT

### 2.1 General description

B-BOP is an imaging instrument, providing diffraction-limited imaging simultaneously in total power mode, i.e. measuring the total flux of the sky, and in a polarimetric mode, where it records signals that are linked to the linear polarized fraction and linear polarization angle of the incoming flux.

Indeed, unlike most polarimetric imagers, polarization is not measured using an optical device that splits the incoming light according to its polarization state. Rather B-BOP's detector arrays are composed of two types of pixels where, in each pixel, we record the power imbalance between two orthogonal grids of absorbers, with a rotation of 45° between the two types of pixels. Pixels are arranged in a checkerboard manner in the focal plane, and the pixel sizes are adjusted so that the spatial sampling frequency of each type of pixel is well within the Nyquist criterion. Full spatial sampling is recovered by scanning the sky.

B-BOP carries 3 imaging channels, or bands, centered at 70 μm for band 1, 200 μm for band 2 and 350 μm for band 3. Each channel has a resolving power ($R = \lambda/\Delta\lambda$) of order 2-3. B-BOP has a common entrance optics for the three bands. Two dichroic filters define the three optical paths that correspond to the three different bands. Different magnifications are then used to get the same image sampling in the three bands (0.6 F.Lambda). The Field of View is almost identical for the three arrays (3.2' x 3.2' to 3.7' x 3.7'). A 4K Calibration Unit (with an adjustable linearly polarized source) will enable the critical photometry and polarimetry calibrations.

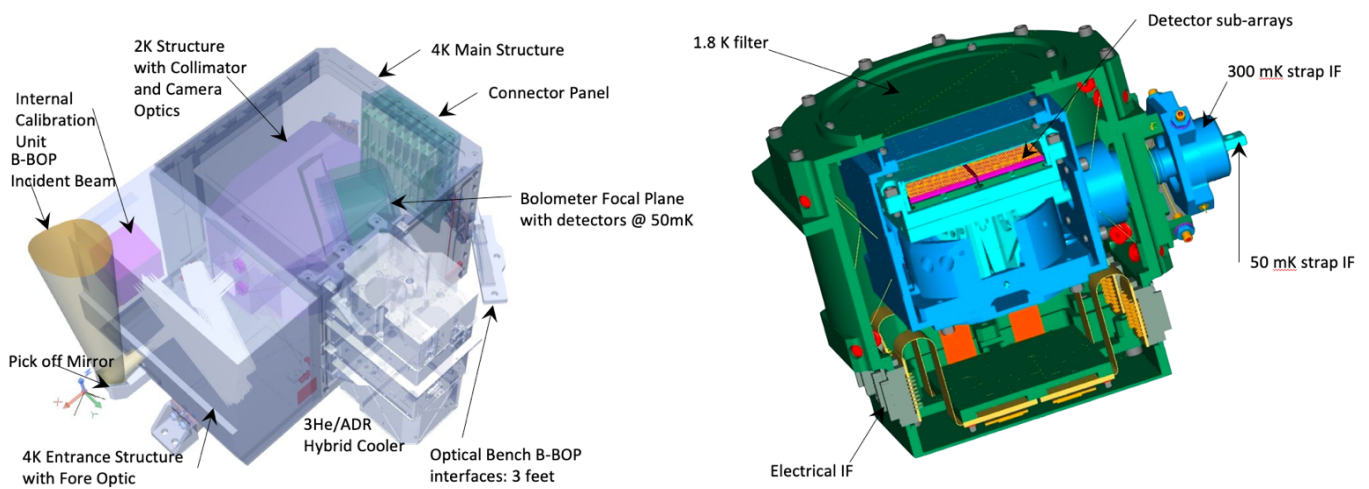

Figure 1. Left : 3D view of the B-BOP instrument (Focal Plane Unit). Right : the Bolometer Focal Plane.

B-BOP consists of three main elements:

**The Focal Plane Unit (FPU)** which sits on the Instruments Mounting Structure (IMS), next to the SMI and FGS focal plane units, and on the opposite side of the SAFARI focal plane units. This FPU is itself made of subsystems:

- The Bolometer Focal Plane (BFP) that hosts the detectors and their readout circuits
- The hybrid Cryo-Cooler (HCO) that delivers the 50 mK and 300 mK temperature levels to the BFP. The HCO is mounted on the 4 K structure.
- The Camera Optics, enclosed in a 2 K mechanical structure, where the beam is split in 3 bands that are imaged on the BFP.
- The 2 K mechanical structure, on which the BFP is attached and which is itself within the 4 K mechanical structure, thermally insulated from it.
- The Entrance Optics, mounted in the 4 K mechanical structure, that is used to compensate for induced polarization by the outer optical elements as well as ensure proper protection from stray light.
- The Internal Calibration Unit that is used to estimate and monitor the amount of instrument polarization.
- The pick-off mirror that redirects part of the SPICA telescope beam in the B-BOP instrument.
- The 4 K mechanical structure.

**The Cold Front-End Electronic**s which is a relay element that serves to carry the signals from the bolometers to the electronics on the service module (SVM). It is in principle located either at the telescope shield level or at the 1st V-Groove level.

**The Warm Electronics**, on the SVM performs all the control and readout functions of the instrument. It is composed of three elements:

- The Warm Front-End Electronics (WFEE) that controls the detectors (through the application of biases) and reads them (with clocks that control the multiplexing system).
- The SubSystems Control Unit (SSCU) is dedicated to the control of the Hybrid Cooler and the Internal Calibration System, as well as to the readout of all temperature information on the FPU.
- The Data Processing Unit (DPU) that processes uplink commands for the control of the instrument subsystems and packages the readout and housekeeping information for downlink.

## 2.2 Optical design

The optical chain is composed of two main systems, the Fore Optics and the Imager Optics. This separation allows distributing functions of the optics in separate elements which will later make for easier distribution of the manufacturing, verification, validation and integration work.

The fore-optics functions are the following:

- Collect from the main telescope beam the B-BOP field of view. This is the role of the Pick-Off mirror which is the only element of B-BOP that is located outside the 4 K housing.
- Implement cold field stops to provide control over straylight propagation inside the instrument.
- Compensate as much as possible polarization induced by the telescope and POM.
- Accommodate the Internal Calibration Unit at a pupil image location.
- Create an intermediate focal plane IFP at the entrance of the imager optics.

The fore-optics is mounted on the 4K structure and thermalized at this temperature level. These components must be kept at a temperature below 5.5K in order to prevent them from becoming the dominant source of noise for our detectors.

The role of the imager optics is to image the field of view on the B-BOP detectors. B-BOP observes a common FoV in three bands simultaneously and thus the imager has in effect 3 different optical paths. In order to do this, the imager optics contains two dichroic mirrors that allow separation of the different wavelength ranges and the re-direction of the corresponding light beams to the proper optical paths.

The imager optics is enclosed in a light-tight mechanical structure that is maintained at the 1.8 K level so has to contribute as little as possible to the background falling on the detectors

The imager optics consists first of a collimator that provides a second image of the pupil of the telescope at which we can implement another cold stop. This cold stop is coincident with the first dichroic mirror and signals the beginning of the three different imagers.

### 2.3 The focal plane unit

The Bolometer Focal Plane integrates the 6 detectors sub-arrays. The Russian-dolls optical and mechanical system scheme is chosen in order to control and provide the requested nearby environment to the detectors considering their very high sensitivity and the very stringent requirement on the noise.

The 6 individual detector arrays are bolted on a common plate considered as the mechanical reference of the focal plane. This common plate is supported and insulated from the 300mK stage by a set of Kevlar wires with pulleys and capstans similar to the Kevlar suspension system well qualified for the PACS[3] instrument of the Herschel observatory.

### 2.4 The hybrid cryo-cooler

The B-BOP bolometers require an operating temperature of 50 mK. This is obviously not delivered by the cryogenic facility of the spacecraft that delivers only 4.8 K and 1.8K. Therefore, to bridge the last gap, the instrument has its own cryogenic component in the form of a hybrid cryo-cooler (HCO). As the present time, this hybrid cryo-cooler is the same as the one found in the SAFARI instrument with the notable exception that the amount of magnetic field shielding is reduced both to save mass in the Payload Module (PLM) and because the B-BOP detectors are much less sensitive to magnetic perturbations. In the course of the instrument development the two HCOs may diverge but we note that the performance of this line of HCO can be adjusted significantly without requiring significant design changes.

The HCO is called "hybrid" because it is made of two stages. At the first stage we find a helium sorption cooler that generates a 300 mK level when liquid helium is evaporated (using pumping charcoals to force helium evaporation therefore drawing heat from the system). At the second stage we use an adiabatic demagnetisation refrigerator. This uses the fact that relaxation of the spins in a paramagnetic salt from a fully ordered state to a fully disordered state is also a process that draws heat.

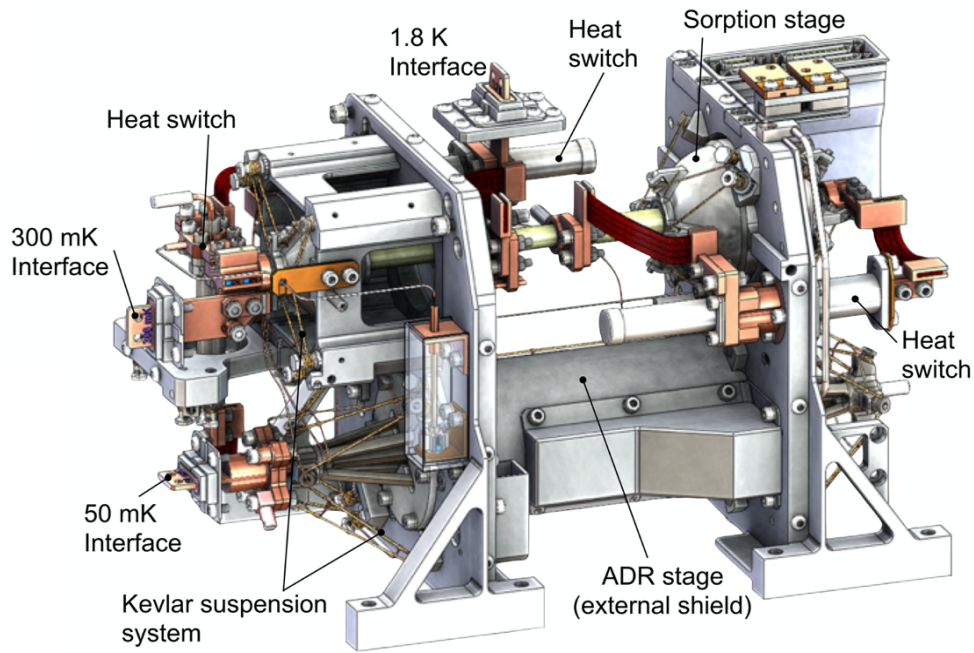

Figure 2. Sketch of the hybrid cryo-cooler for B-BOP.

The Sorption stage is built of two main elements: an evaporator in which liquid Helium 3 is condensed and then pumped to keep it evaporating, and a sorption pump where the Helium gas is captured during the evaporation process. Both are connected by a pumping line where Helium gas can circulate from one element to the other. The pump uses active charcoal to capture the Helium. The cooler uses heat switches to maintain either the pump or the evaporator, depending on the phase of the process, connected to a heat sink. For SPICA the heat sink is the 1.8 K level for the evaporator side, and the 4.5 K level for the pump side.

The sorption stage is in operation while there is liquid helium in the evaporator. In that case the evaporator is thermally isolated from all systems except those one the cold side (detector support, ADR cooler, see below). When all the Helium has been vaporized the sorption cooler must be recycled. Recycling of the sorption cooler requires connecting the evaporator to the 1.8 K level (and thus disconnecting it from the cold side), while heating the pump to ~30 K to vaporize the gas. This gas will flow to the cold evaporator and liquefy. At the end of this process the pump is connected to the 4.8 K stage to get rid of its heat and the cycle can begin again.

The principle of this stage is to cycle between a strong magnetic field phase where this field is used to align the spins in a paramagnetic salt, and a phase where the magnetic field is slowly decreased so that the spins can turn back to their random orientation, albeit at a controlled rate. The first phase sees entropy decreasing and thus leads to heat release. This has to happen while connected to the 300 mK level so as to get rid of this heat above the ADR stage. The second phase proceeds which the ADR cooler is isolated from the warmer stages and as the return to disorded stage requires heat, it will cool the paramagnetic salt to a temperature lower than 300 mK. One of the very interesting properties of the ADR stage is that one can use the rate at which one decreases the magnetic field to control the temperature of the salt, and adjust it to the thermal load it receives from the detectors.

When the magnetic field reaches 0T the cooling power of the ADR stage has been exhausted and the ADR stage must be recycled.

## 2.5 The detectors

The detectors technology[1] derives from the *all-silicon* bolometers arrays developed for Herschel/PACS in the 2000's, but with a totally different design. Here, the light absorption is made by dipole networks covering ~ 85% of the pixel[2]. The dipoles of same orientation are linked together by a thermally insulated silicon spiral structure, interleaved with a second spiral carrying the orthogonal dipoles. The two spirals are doped with donors [P] and acceptors [B] in different densities and act as a very sensitive thermometric sensor below 100 mK. This thermal sensor is suspended by microscopic copper posts (10 μm diameter) above a reflector defining a quarter wave cavity (11 μm for the 70 μm band, and 24 μm for the 200 μm band), that ensure a very large absorption efficiency (>90%).

The electromagnetic energy of light is transferred to heat in the suspended structure by matching the dipole resistance to the *local* vacuum impedance. The temperature elevation changes the resistance of the thermal sensors and is converted into signals. To reach the targeted wavelength absorption, the cavity is partially filled with Silicon oxide. The section of the silicon spirals is 1.5μm x 4 μm, and the etching depth between spirals and $SiO_2$ substrate is 2 μm for a ~1.3 mm spiral length. Every other pixel in the two directions of a chessboard pattern array has dipoles rotated by 45° (see figure 3) in order to access to the Stokes Parameters for each beam in the sky, considering that the Airy disc area is equal to ~ 12 pixels in the focal plane (Nyquist spatial sampling).

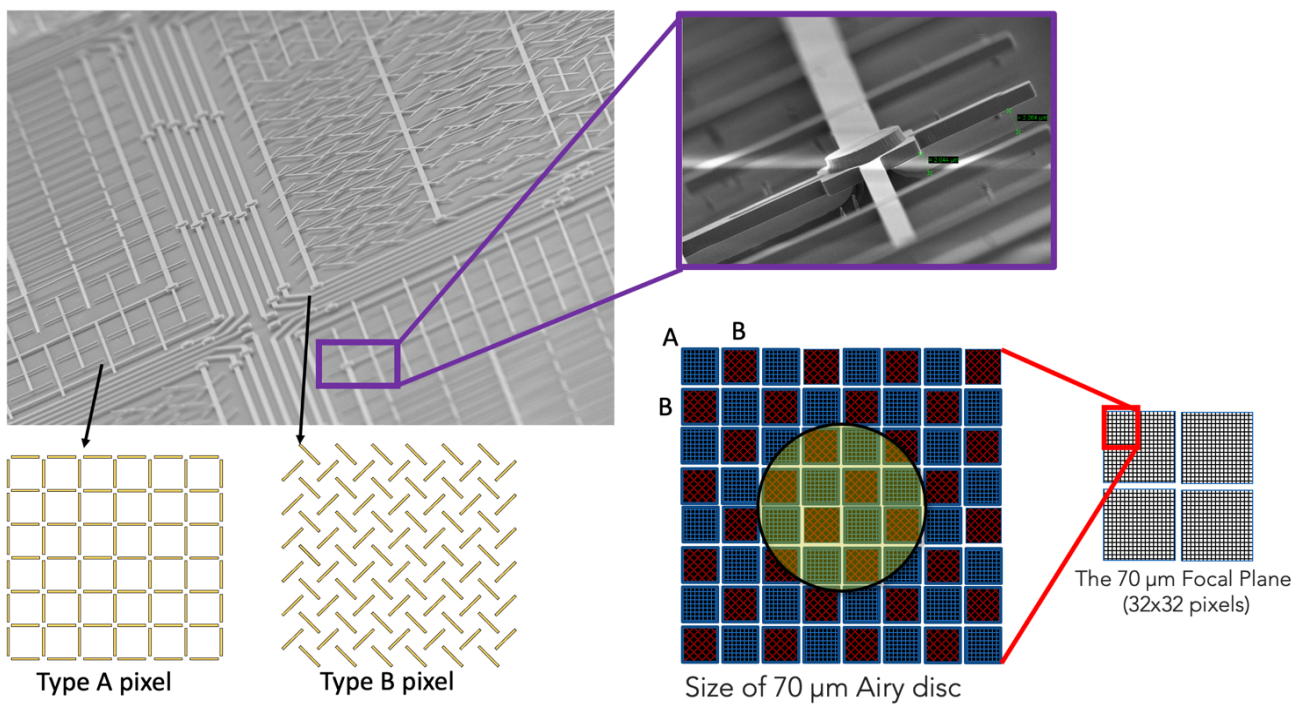

Figure 3. Details of the B-BOP pixels. Top left and right are SEM images of these pixels, left shows the dipoles supported by the silicon spirals, the right one shows the detail of a copper nail supporting the spiral. Below are sketches showing the two types of pixels carrying two dipole patterns.

The basic principle starts with a fully symmetrical Wheatstone bridge that is implemented in the pixel, as shown on Figure 4. This electrical circuit contains fixed load resistors (Rc) which are located on each side of the pixel and thermally connected to the 50 mK bath, and variable resistors (RH and RV). The H and V refers to Horizontal and Vertical. Indeed, these variable resistors correspond to the cyan, red, blue and yellow "spirals" in the pixel (right-hand panel of figure 4) which are doped silicon beams that carry a grid of either horizontal or vertical dipoles. The red and yellow spirals carry horizontal dipoles and thus each constitutes a RH resistor, while the cyan and blue spirals carry vertical dipoles and constitute the RV resistors.

Because each spiral carries dipoles oriented in a single direction, it couples to the incoming light in a way that is dependent on the polarization fraction, p, of this light and of its polarization angle, ψ.

On each pixel we measure two quantities, indicated on the left of figure 4. Vamp is linked to the total amplitude of the incoming power, while Vimb is linked to the imbalance between the power absorbed by the vertical and horizontal grids and thus is a function of p and ψ. To disambiguate the two, we use a second type of pixel where we have the same doped silicon spirals but where the dipoles have been rotated by 45°. So instead of measuring the imbalance of the power absorbed by a grid at 0° with respect to a grid at 90°, we measure the imbalance of the power absorbed by a grid at +45° with respect to a grid at -45°. A detector array is thus made of a succession of these two types of pixels, arranged in a checkerboard manner.

To adjust the wavelength at which the peak of the pixel absorption occurs we play on two geometrical parameters. The first one is the size of the absorbing grid cell (defined by the width and length of the dipoles and their spacing) and the second one is the depth of the quarter-wave cavity that is built below the absorbing grid (and above the readout circuit). This is sufficient to manufacture the band 1 and band 2 detectors (70 and 200 μm) however it is not possible to use the same process for the band 3 (350 μm) as the quarter wave cavity becomes too deep. In that case the principle we use is to deposit on top of the absorbing grid a layer of silicon which is modifying the absorption profile of the pixel shifts the peak toward longer wavelength[4].

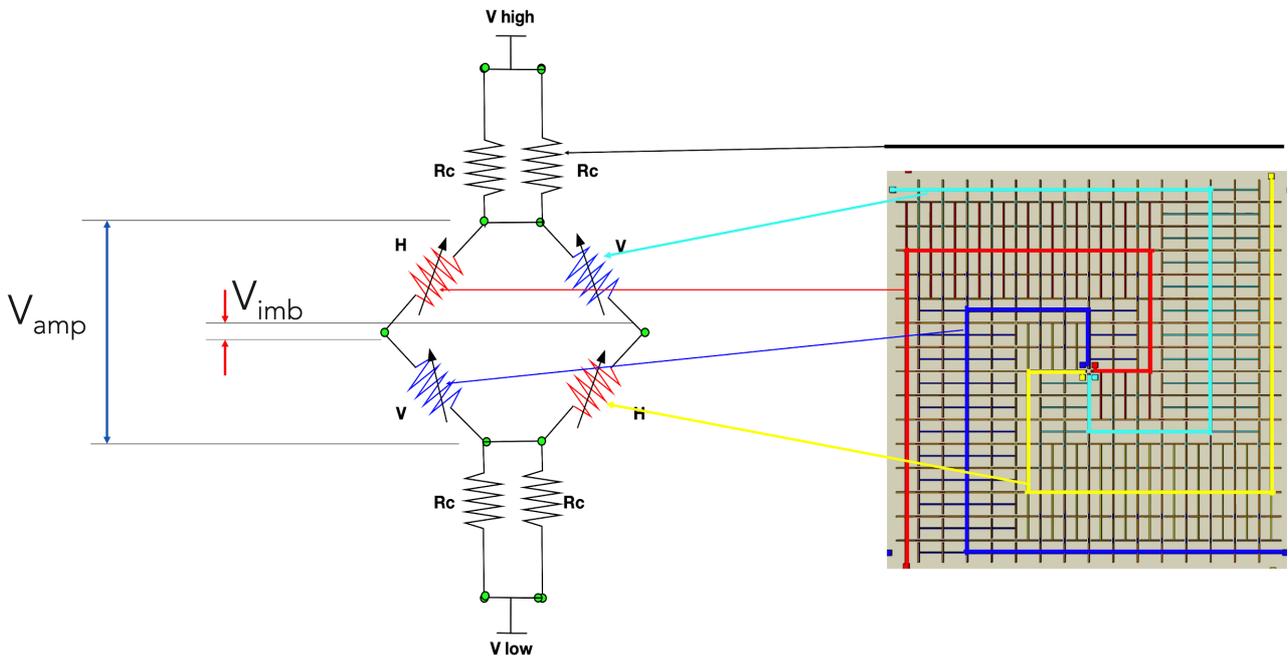

Figure 4. The principle of B-BOP's pixel with, on the right a representation of the physical implementation of the pixel and on the left the corresponding electrical circuit.

The readout circuit is implemented in the ASIC "Eclipse" on which the sensitive part of the detector has been built. This ASIC is used to set the bias at the Wheatstone bridge ($V_{high}$ and $V_{low}$) as well as a series of other biases that are used to open and close the readout circuit and thus achieve the 16x (bands 1 and 2) or 8x (band 3) multiplexing.

The readout frequency of 40 Hz corresponds to the frequency at which the whole array is read once. This means that for a multiplexing factor of 16, we read one column of 16 pixels every 1/(40x16) second, so that the 16 columns of an array have been read in 1/40 seconds.

As explained above "reading" a pixel is measuring (at least) 2 bias differences per pixel ($V_{amp}$ and $V_{imb}$). During the development phase of the detector we actually intend to record more than these two biases and thus the readout circuit is built to allow such flexibility. For flight operations, the readout sequence will be selected as the one that most efficiently allows accessing the total power and polarization properties, and it will likely be the only sequence implemented.

While the readout circuit is implemented just below the detection layer, a buffer unit is located on the BFP at 2 K to amplify the signals so that they reach the CFEE.

# 3. OBSERVATIONS, CALIBRATION, AND EXPECTED PERFORMANCES

## 3.1 Observing modes

B-BOP is an imaging polarimeter that will essentially perform large maps, significantly larger than its intrinsic field of view, even though "small maps", for point sources are possible. It will do so by scanning the sky along a rectangular grid in so-called "boustrophedon" mode as is shown on Figure 3. In this mode we use the spacecraft to perform long regular slews, called scan legs, in one direction, followed by a small step in the perpendicular direction. The length of a scan leg will typically range between some tens of arcminutes (e.g. when mapping individual objects and maximising the spatial coverage) to a few degrees (e.g. when mapping complete interstellar clouds or sections for the Galactic plane).

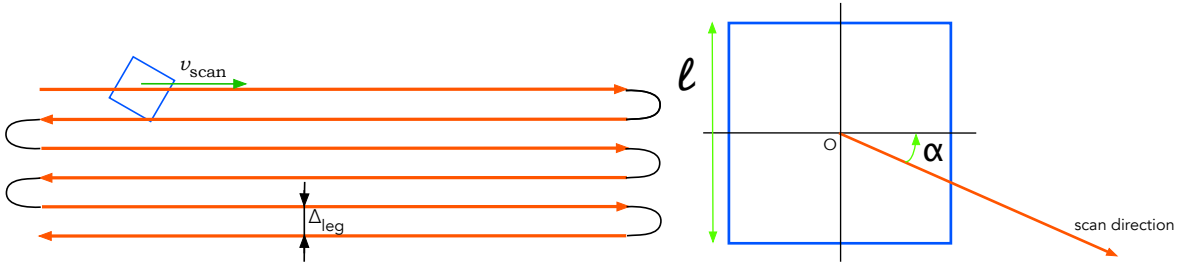

Figure 5: On the left, the Boustrophedon scanning mode, where the telescope scans the sky in parallel lines, with curved trajectories in black to go from one line to the next. The angular separation between legs is noted Δleg. The blue square indicates an instantaneous position of the detector during the scan. On the right, the blue square represents again the array we are considering for the observation. The black lines are the reference axes for this array which is made of square pixels (these will be spacecraft axes eventually). The red arrow represents the scan direction and makes an angle α with respect to one of the main axis. Symmetry considerations show that α ∈ [0,π/4] covers all possible scanning strategies. We have also indicated the value of another parameter: $l$ the array side length.

The separation between two scan legs is made such that there is an overlap in the area covered by the back and forth scan legs. The amplitude of this overlap impacts the time needed to cover a given portion of the sky, as well as the sensitivity of the final map. It is quite likely as well that the B-BOP bolometers will present a component of low-frequency (LF) noise which is generally handled by multiplying the number of times a given sky pixel is covered by detector pixels. This can either lead to decreasing the step between two scan legs (i.e. increasing the overlap region), or implementing a mode where the observation is done twice (or more) with different scan leg directions. For Herschel, both bolometer instruments (PACS & SPIRE) were systematically using two orthogonal scan directions per observations.

A further parameter of the observing mode is the scanning speed, i.e. the speed at which the spacecraft is commanded to move along the scan leg. With detectors such as those of B-BOP, the allowed scan speeds are bracketed by two conditions. At the high end, we need to limit the scan speed because bolometers have a finite time constant (i.e. they adjust to a changing illumination in a finite time): when the scan speed is too high, the bolometer signal cannot adjust as fast as the illumination and we observe a blurring effect, or an increase of the point spread function (PSF) in the direction of the scan. This in turns increases the level of the confusion noise, which we want to avoid. At the low end, we are falling back on LF noise limitations: one way to consider this noise's behaviour is as a drift of the signal level, independent of the illumination. The principles used to "filter" data from this noise component often rest on the assumption that the frequency domain in which LF noise dominates is different from that where the sky signal dominates, the latter being defined by the modulation frequency of the sky signal. In the scanning case, we modulate the signal by scanning and thus we must make sure that the scanning speed is large enough that for any pixel, modulation of its illumination (obtained by scanning the sky scene) is done at a higher frequency than the characteristic LF noise frequency. Based on preliminary studies and on our experience with the PACS detectors, we estimate the scanning speed for B-BOP at 20 ''/s.

## 3.2 Calibration

Being an imaging photometer and polarimeter, B-BOP will require calibration of a number of elements that intervene in the transformation of a sky signal into an output signal from the detectors, back into a reconstructed sky signal. A large fraction of the calibration information will come from astronomical source observations, or trend analysis from the instrument behaviour in operations. A smaller fraction will come from ground-based calibrations. However, as polarization is a complex and still underdeveloped aspect of astrophysical investigations, there is both a need to precisely understand the behaviour of our instrument and a lack of well-known astrophysical calibrators.

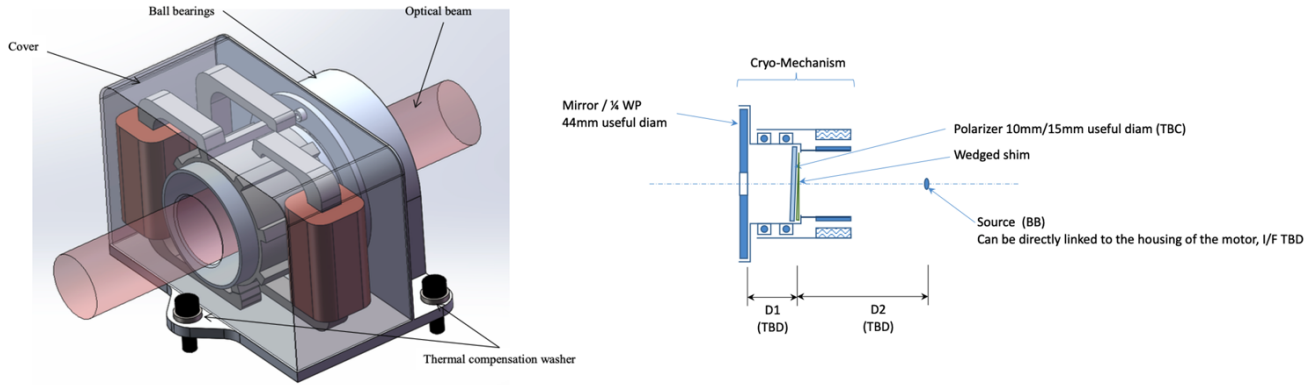

Figure 5. Rendering of the conceptual design for the cryo-mechanism of the calibration unit.

The internal calibration unit (ICU) role is to both determine and monitor the level of instrumental polarization and as well as calibrate the response of our detectors to polarization parameters (polarization fraction and angle). For this reason, the calibration source is able to generate a fixed FIR power, polarized at a level comparable to the levels we will observe in the sky (i.e. of the order of 10%), with an adjustable polarization angle.

**Operating principle**

Since observing with B-BOP relies on no moving part either in the instrument or in the telescope, and since the pointing constraints on the spacecraft placed in order to maintain the telescope below 8 K are rather high, we do not expect that there will be short-term variations in parameters that affect the level of instrumental polarization (where "instrumental" here includes the telescope as well) or the performance of the detectors. Therefore, the operation of the ICU will likely be an activity performed at the beginning of an observing cycle (after cooler recycling) and not repeated afterward.

In the optical chain the ICU is positioned in the entrance optics, behind a mirror that is at an image position of the telescope pupil. Therefore, we use the central obscuration of the telescope to remove the central part of this mirror so that light from the calibration source can be injected in the optical path.

During a calibration sequence, the spacecraft will be pointed to a predetermined position in the sky (chosen to avoid sources with polarization and strong spatial gradients, the calibration source will be turned on and the rotation of the polarization angle will be achieved by rotating the ICU using a cryo-mechanism that will rotate a calibration device located in front of the calibration source.

The mechanical assembly of the ICU effectively consists mostly in the cryogenic mechanism that will be used to achieve rotation of the polarizing device. The calibration source itself will be fixed, possibly attached to the housing of the cryo-mechanism while the polarizing device and the mirror at the pupil image location will be attached to the mechanism. It is hollow so that light from the source can be injected in B-BOP's optical path even if this light source is located behind the cryo-mechanism. On the outside of the rotating cylinder one can see the "teeth" which are used to define the fixed orientations that we will use for calibration. Indeed it is important to note that the calibration sequence will not consist in continuously rotating the ICU but rather in stepping through a series of well-defined positions that sample appropriately the angular range to be calibrated (in that respect is it unlikely that these teeth will be regularly distributed as shown here are some are effectively redundant). This rotating cylinder will hold the polarizing device, currently foreseen as a silicon wafer with parallel edges, mounted with a slight angle, as well as the mirror that belongs to the entrance optics.

### 3.3 Estimated performances

The following table presents the estimated performances of the B-BOP instrument on SPICA, based on preliminary measurements made in the lab (variation of the impedance of the bolometers with temperature, noise, pixel absorption).

All sensitivities are for 5σ, 10h mapping of a 1 square degree area. For polarization, 5σ refers to the polarization fraction p (p/σp=5). Only internal observation overheads are included, stray light and confusion limit are not included.

|  | Band 1 | Band 2 | Band 3 |
|---|---|---|---|
| Band center | 70 µm | 200 µm | 350 µm |
| Band edges | 52—88 µm | 135—225µm | 280—420µm |
| # of pixels | 4 x 16 x 16 | 16 x 16 | 8 x 8 |
| Pixel size | 5" x 5" | 14" x 14" | 28" x 28" |
| Band centre FWHM | 7.6" | 21.7" | 37.9" |
| Astrophysical background surface brightness | 4.75 MJy/sr | 3.64 MJy/sr | 1.87 MJy/sr |
| Point Source sensitivity (unpolarised) | 0.28 mJy | 0.82 mJy | 1.55 mJy |
| Point Source sensitivity in Stokes (Q,U) at 5% polarization level | 8.0 mJy | 23.2 mJy | 43.9 mJy |
| Surface brightness sensitivity (unpolarised) | 0.12 MJy/sr | 0.043 MJy/sr | 0.023 MJy/sr |
| Sensitivity to map Stokes (Q,U) at 5% polarization level | 3.4 MJy/sr | 1.25 MJy/sr | 0.64 MJy/sr |

Table 1. Expected performances for B-BOP on the SPICA telescope.

## 4. CONCLUSION

We present in this paper the status of the development of the B-BOP instrument on SPICA. B-BOP is an imaging polarimeter that contains three channels of polarization sensitive ultra-sensitive bolometers. Despite the cancellation of the SPICA mission during the phase A, the work continues with plans to test the B-BOP detectors on a ground-based facility or onboard a balloon-borne experiment.

This work was partially funded by CNES, the European Space Agency (contract No. 4000130941/20/NL/IB/ig) and Labex FOCUS (ANR-11-LABX-0013).

*vincent.reveret@cea.fr